\documentstyle[12pt]{article}

\renewcommand{\thefootnote}{\fnsymbol{footnote}}
\newcommand{\vs}[1]{\vspace{#1 mm}}

\begin{document}
\baselineskip 5mm

\newcommand{\ov}{\overline}

\begin{titlepage}
\setcounter{page}{0}
\begin{flushright}
KEK-TH 841\\
September 2002\\
\end{flushright}

\vs{6}
\begin{center}
{\LARGE  Fuzzy Algebrae of  \\ the General  
K\"ahler Coset Space $G/S\otimes{U(1)}^k$ }

\vs{6}
{\large
Shogo Aoyama$^{1}$\footnote{e-mail: spsaoya@ipc.shizuoka.ac.jp}\ \ \ \ and 
\ \ Takahiro Masuda$^{2}$\footnote{e-mail: stmasud@post.kek.jp} 
}\\
\vs{6}
{\em ${}^1$ Department of Physics, Shizuoka University \\
          Ohya 836, Shizuoka, Japan  \\
 \   \\         
      ${}^2$ High Energy Accelerator Research Organization (KEK),\\ 
               Tsukuba, Ibaraki 305-0801, Japan}
\end{center}
\vs{20}

\centerline{{\bf{Abstract}}}
\vs{3}

We study the fuzzy structure of the general K\"ahler coset space $G/S\otimes{U(1)}^k$  deformed by the Fedosov formalism. It is shown that the Killing potentials satisfy 
 the fuzzy algebrae working in the Darboux coordinates.

\vs{65}
\begin{flushleft}
\end{flushleft}

\end{titlepage}
\newpage

\renewcommand{\thefootnote}{\arabic{footnote}}
\setcounter{footnote}{0}

\newpage

The concept of non-commutative space-time is age-old in physics, for smooth space-time is expected to get modified at the planck scale due to quantum corrections. The discovery of noncommutativity of space-time in the M-theory and the string theory shed new light on this old theme\cite{0}. It attracted much attention  and began to be studied
with more concrete motivations. The foundation of the study got firm owing to works by many people. (See, for instance, \cite{1} for the recent review and references therein.) Non-commutativity in curved space-time is a much involved subject in contrast with that in flat space-time. There we expect to uncover  rich physical structures  due to non-commutative geometry. One of the methods to explore non-commutative geometry is quantum deformation of manifolds. For the symplectic manifold Fedosov gave a general formalism to this end\cite{2}. It provides  a definite way to make quantum deformation of manifolds and to define an associative non-commutative $\star$ product on deformed manifolds.
\vspace{1cm}

In refs \cite{3,4} we have studied the fuzzy algebrae for the K\"ahler coset space $G/H$ 
 deformed by the Fedosov formalism. In ref. \cite{3} it was shown that   the Killing potentials \cite{5,6} satisfy 
in the holomorphic coordinates the fuzzy algebrae 
\begin{eqnarray}
 [ M^A(z,\ov z), M^B(z,\ov z) ]_\star
 &=& -i(c_1 \hbar + c_3\hbar^3 + c_5\hbar^5 + \cdots)\sum_{A=1}^{dim\ G}f^{ABC} M^C(z,\ov z),  \label{eq 0} \\
\sum_{A=1}^{dim\ G} M^A(z,\ov z)\star M^A(z,\ov z) &=& c_0  + c_2\hbar^2 + c_4\hbar^4 + \cdots\ ,  \label{eq 00}
\end{eqnarray}
when the coset space is irreducible. 
The coefficients $c_0,c_1,c_2, \cdots $ are numerical constants. 
In ref. \cite{4}  we found these coefficients as $c_i = 0$ for $i\ge 3$,  if we work in a particular set of the Darboux coordinates. 
In this letter  we will show that  this finding in ref. \cite{4} is true for  the general K\"ahler coset space which is not necessarily irreducible.  The irreducibility of the K\"ahler coset space was essential for the demonstration in ref. \cite{4} . To generalize it we have to endeaver a new strategy without recourse to the irreducibility.

\vspace{1cm}

We shall start with a brief summary on the general K\"ahler coset space\cite{7,8}. According to the the theorem by Borel\cite{9} a coset space $G/H$ is k\"ahlerian if the unbroken subgroup $H$ is such that ${\rm rank}\ G = {\rm rank}\ H$ and
\begin{eqnarray}
H = S\otimes \{U(1)\}^k, \quad\quad\quad k = 1,2,\cdots\cdots \le {\rm rank}\ H.
  \nonumber
\end{eqnarray}
 We decompose the generators $T^A$ of $G$ as
\begin{eqnarray}
   \{ T^A\} &=& \{ X^a, S^I, Q^\mu \},  \label{gene} \\
     a & = & 1,2,\cdots,2N(= {\rm dim}\ G - {\rm dim}\ H),\nonumber\\
     I & = & 1,2,\cdots,{\rm dim}\ S (={\rm dim}\ H - k), \nonumber \\
     \mu & = & 1,2,\cdots,k,   \nonumber
\end{eqnarray}
in which $S^I$ and $Q^\mu$ are generators of the unbroken subgroups $S$ and $U(1)$s' respectively, while $X^a$  broken generators. Let us define a central charge as 
\begin{eqnarray}
Y = \sum_{\mu=1}^k v^\mu Q^\mu.   \nonumber
\end{eqnarray}
With a generic choice of the real coefficients $v^\mu$ all the broken generators $X^a$ have non-vanishing $Y$-charges. Then they can be split into two parts: 
generators
$X^{\ov i}$ with positive charge and $X^i$ with negative charge such that
\begin{eqnarray}
 [Y,X^i] &=& -y_i(v)X^i,  \quad\quad\quad
 [Y,X^{\ov i}] \ \ = \ \  y_i(v)X^{\ov i},   \label{y}
\end{eqnarray}
with no sum over the indices $i, \ov i = 1,2,\cdots,N$. Accordingly we further decompose the generators of $G$ in (\ref{gene})  as
\begin{eqnarray}
\{ T^A\} = \{ X^{\ov i},X^i, S^I, Q^\mu \}.
\end{eqnarray}
If the broken generators $X^i$ are in an irreducible representation under the unbroken subgroup $S$, then the $Y$-charges $y_i(v)$ are all the same. Otherwise $X^i$ are decomposed into irreducible sets, each of which may have a different $Y$-charge, due to the Schur's lemma. This argument defines the irreducibility of the K\"ahler coset space $G/H$. For the irreducible K\"ahler coset space $G/H$, the number of the $U(1)$ generators $Q^\mu$ is necessarily  one, that is, $Y\propto Q$. Then the group $G$ satisfies the Lie-algebra in the form 
\begin{eqnarray}
[ X^i, X^{\ov j}] &=&  (\Gamma^I)^{i\ov j} H^I +
 \delta^{i \ov j} Y,  \quad\quad  [ X^i, X^j ] \ =\  0,  \nonumber \\
 \quad [ X^i, H^I ] &=& (\Gamma^I)^{i\ov j} X^j\quad\quad [X^i, Y]\ =\  X^i, \quad c.c..   \label{Lie}
\end{eqnarray}
For the general K\"ahler coset space the Lie-algebra does not take such a simple form.
\vspace{2cm}

We next summarize the geometrical properties which hold for the K\"ahler coset space 
in general\cite{5}. The line-element is given by the metric of the type $(1,1)$:
\begin{eqnarray}
ds^2 = g_{\alpha\ov \beta} dz^\alpha dz^{\ov \beta}.  \nonumber
\end{eqnarray}
The K\"ahler form 
\begin{eqnarray}
d\omega= g_{\alpha\ov \beta} dz^\alpha \wedge dz^{\ov \beta},  \label{Kaform}
\end{eqnarray}
is closed. It implies that 
the metric $g_{\alpha\ov \beta}$ is derived from a real K\"ahler potential $K(z,\ov z)$ as 
\begin{eqnarray}
g_{\alpha\ov \beta} =  K_{,\alpha\ov \beta}.
\end{eqnarray}
Since the K\"ahler manifold is a coset space, 
there exists a set of holomorphic Killing vectors 
$R^{A\alpha}(z), A = 1,2,\cdots,{\rm dim}\ G$,  
which represent the isometry $G$
\begin{eqnarray} 
 {\cal L_{R^A}}R^{B\alpha} = R^{A\beta} R^{B\alpha}_{\ \  ,\beta} - R^{B\ \beta} R^{A\alpha}_{\ \ ,\beta} 
  \ \ = \ \ f^{ABC}R^{C\alpha}.   \label{LM}
\end{eqnarray}
They satisfy the Killing equation
\begin{eqnarray} 
R^A_{\ \alpha,\ov\beta} + 
R^A_{\ \ov\beta,\alpha} \ \ = \ \ 0,     \label{sss}
\end{eqnarray}
with$R^A_{\ \alpha} = g_{\alpha\ov\beta}R^{A\ov\beta}$ and $R^A_{\ \ov\alpha} = g_{\beta\ov\alpha}R^{A\beta}$. From this we may find a set of real Killing potentials $M^A(z,\ov z)$ such that
\begin{eqnarray}
R^A_{\ \alpha} = \ i M^A_{\ ,\alpha} , \quad \quad  
R^A_{\ \ov \alpha} = - i M^A_{\ ,\ov \alpha}.  \label{R}
\end{eqnarray}
By the Lie-variation with respect to $R^A$ the Killing potentials transform as the adjoint representation of $G$
\begin{eqnarray}
{\cal L_{R^A}} M^B \equiv  R^{A\alpha} M^B_{\ ,\alpha} +
            R^{A\ov\alpha}M^B_{\ ,\ov\alpha} 
                     = f^{ABC}M^C. \label{MA}   \label{RR}
\end{eqnarray}
On the other hand the K\"ahler potential transforms as 
\begin{eqnarray}
{\cal L_{R^A}} K & \equiv & K_{,\alpha}R^{A\alpha} + K_{,\ov \alpha}R^{A\ov \alpha}
  \nonumber   \\
   & = & F^A + \ov F^A,  \label{LieK}
\end{eqnarray}
with some  holomorphic functions $F^A(z)$ and their complex conjugates $\bar F^A(\bar  z)$.

\vspace{1cm}

The idea to investigate the fuzzy algebrae of the general K\"ahler coset space is 
to use 
the Killing potentials expressed in the form
\begin{eqnarray}
iM^A = K_{,\alpha}R^{A\alpha} - F^A,  \label{kill}
\end{eqnarray} 
or their complex conjugates\cite{5}.
This relation can be easily shown  by integrating the second equation in (\ref{R}) by $z^{\ov \alpha}$ and using the property (\ref{LieK}). Let us change the local coordinates $(z^\alpha, z^{\ov \alpha})$ to  the ones given by $(q^\alpha,p^\alpha)$ 
\begin{eqnarray}
q^\alpha = z^\alpha, \quad\quad\quad p_\alpha = iK_{,\alpha}.
\end{eqnarray}
Then the  K\"ahler form (\ref{Kaform}) can be put in the form 
\begin{eqnarray}
d\omega = dp_\alpha \wedge dq^\alpha.    \nonumber
\end{eqnarray}
Hence $(q^\alpha,p^\alpha)$ are the Darboux coordinates. 
In terms of them the Killing potentials (\ref{kill}) become 
\begin{eqnarray}
iM^A = p_\alpha R^{A\alpha}(q) - F^A(q).  \label{lll}
\end{eqnarray}
Here we explicitly wrote the arguments of $R^A$ and $F^A$ to make the fact evident that they are holomorphic in $z$. 
As has been discussed in ref. \cite{4}, in the Darboux coordinates the Fedosov $\star$ product for the K\"ahler manifold 
locally   reduces to 
\begin{eqnarray}
 a(p,q)\star b(p,q) =
                 \sum_n {1\over n!} a(p,q)[ {i\hbar \over 2}
                             ({\overrightarrow \partial \over \partial p_\alpha }
                             {\overleftarrow \partial \over \partial  q^\alpha}
                           -  {\overleftarrow\partial \over \partial p_\alpha }
                             {\overrightarrow \partial \over \partial q^\alpha})]^n b(p,q),
                         \label{Moy}
\end{eqnarray}
which is the Moyal product.

\vspace{1cm}

With the Moyal  product (\ref{Moy}) we now examine the fuzzy algebrae (\ref{eq 0}) and (\ref{eq 00}). It is straightforward to show the first one:
\begin{eqnarray}  
 [M^A, M^B ]_\star & =& -i\hbar ( M^A_{\ ,\alpha}R^{B\alpha} - M^B_{\ ,\alpha}R^{A\alpha}) \nonumber  \\
 & = & -i\hbar ( M^A_{\ ,\alpha}R^{B\alpha} + M^A_{\ ,\ov\alpha}R^{B\ov\alpha}) = 
 -i\hbar f^{ABC}M^C
  \nonumber  
\end{eqnarray}
by (\ref{R}) and (\ref{RR}). But the demonstration of the second one is not so simple. Indeed we have
\begin{eqnarray}
M^A\star M^A = M^A M^A + \hbar^2 R^{A\alpha}_{\ \ ,\beta}
R^{A\beta}_{\ \ ,\alpha}.   \label{fuz}
\end{eqnarray}
If the K\"ahler coset space is irreducible, one can show that each  piece in the right hand side is constant, having  explicit formulae for $M^A$ and $R^A$ owing to the simple form of the Lie-algebra (\ref{Lie}). (See ref. \cite{4}.) 
 For the reducible  K\"ahler coset space such formulae are not available. We have to look for a new tactics to show that the right hand side of (\ref{fuz}) is really constant. We have recourse to the theorem: if a real function $f(z,\ov z)$ is invariant by the Lie differentiation
\begin{eqnarray}
{\cal L_{R^A}}f \equiv  R^{A\alpha} f_{\ ,\alpha} +
            R^{A\ov\alpha}f_{\ ,\ov\alpha} 
                     = 0,     \label{kkk}
\end{eqnarray}
$f$ should be constant. It can be shown as follows. (\ref{kkk})
 implies that $i R^{A\alpha} f_{\ ,\alpha}$ are real. Such a real function $f$  can not be other than constant, because $R^{A\alpha}$ are holomorhic and non-linear functions in $z^\alpha$. 
By using this theorem we have 
$$
M^AM^A = const..
$$
We then find that
$$
0 = \{(M^AM^A)_{,\ov \alpha}g^{\alpha\ov \alpha}\}_{,\ov\beta}g^{\beta\ov\beta}
 = 2R^{A\alpha}R^{A \beta}. 
$$
This is a proper identity relation of the holomorphic Killing vectors of the general K\"ahler coset space. 
 Using this identity relation and the Killing equation (\ref{sss}) we manipulate the second 
 term of (\ref{fuz}):
\begin{eqnarray}
R^{A\alpha}_{\ \ ,\beta}R^{A\beta}_{\ \ ,\alpha} &=&
 R^{A\alpha}_{\ \ ;\beta}R^{A\beta}_{\ \ ;\alpha} =
  g^{\alpha\ov\gamma}g^{\beta\ov\delta}R^A_{\ \ov\gamma, \beta}R^A_{\ \ov\delta,\alpha}     \nonumber \\
 &=&  R^{A\ov\alpha}_{\ \ ;\ov\beta}R^{A\ov\beta}_{\ \ ;\ov\alpha}  
  =  R^{A\ov\alpha}_{\ \ ,\ov\beta}R^{A\ov\beta}_{\ \ ,\ov\alpha}. \nonumber 
\end{eqnarray}
 This implies that the holomorphic quantity 
$R^{A\alpha}_{\ \ ,\beta}R^{A\beta}_{\ \ ,\alpha}$ is real, so that it should be constant. Thus our claim has been shown.

\vspace{1cm}

It is instructive to concretize our arguments for the reducible K\"ahler coset space 
$SU(3)/U(1)^2$. We take the Lie-algebrae of $SU(3)$ in the form 
$$
  [T^j_i,\ T^l_k] = \delta_k^j T^l_i - \delta^l_i T^j_k,
$$
by identifying the generators $ \{T^A\} = \{ X^{\ov 1},X^{\ov 2},X^{\ov 3}, X^1,X^2,X^3, Q,Q' \}$ to be 

\begin{center}
\begin{tabular}{ll}
$X^1 = T^2_1,$&$X^{\ov 1} = T^1_2,$\\
$X^2 = T^3_1,$&$X^{\ov 2} = T^1_3,$\\
$X^3 = T^3_2,$&$X^{\ov 3} = T^2_3,$\\
$Q = \textstyle{1\over \sqrt 2}(T^1_1 - T^2_2),$&$Q'= -\textstyle{\sqrt {3\over 2}}T^3_3.$\\
\end{tabular}
\end{center}
Then we find 
the Killing vectors as
\begin{center}
\begin{tabular}{lll}
$R^{\bar{1}1}=i,$&$R^{11}=-i(z^1)^2,$&$R^{Q1}=-\sqrt{2}iz^1,$\\
$R^{\bar{2}1}=0,$&$R^{21}=-iz^1(z^2+{1\over 2}z^1z^3),$&$R^{Q'1}=0,$\\
$R^{\bar{3}1}=0,$&$R^{31}=i(z^2+{1\over 2}z^1z^3),$& \\
$R^{\bar{1}2}=-{i\over 2}z^3,$&$R^{12}=-{i\over 2}z^1(z^2+{1\over 2}z^1z^3),$&$
R^{Q2}=-{i\over \sqrt 2}z^2,$\\
$R^{\bar{2}2}=i,$&$R^{22}=-i[(z^2)^2+{1\over 4}(z^1z^3)^2],$&$R^{Q'2}=-
\sqrt{3\over 2}iz^2,$\\
$R^{\bar{3}2}={i\over 2}z^1,$&$R^{32}=-{i\over 2}z^3(
z^2-{1\over 2}z^1z^3),$& \\
$R^{\bar{1}3}=0,$&$R^{13}=-i(z^2-{1\over 2}z^1z^3),$&$R^{Q3}={i\over \sqrt{2}}z^3,$\\
$R^{\bar{2}3}=0,$&$R^{23}=-iz^3(z^2-{1\over 2}z^1z^3),$&$R^{Q'3}=-
\sqrt{3\over 2}iz^3,$\\
$R^{\bar{3}3}=i,$&$R^{33}=-i(z^3)^2,$& 
\end{tabular}
\end{center}
which satisfy the relation (\ref{LM}). The K\"ahler potential can be calculated according to ref. ref. \cite{7}:
$$
K(z,\bar{z}) = u\log f_1(z,\bar{z})
+ u'\log f_2(z,\bar{z}), 
$$
with 
\begin{eqnarray}
f_1=&1+z^1z^{\bar 1}+
(z^2+{z^1z^3\over 2})(z^{\bar 2}+{z^{\bar{1}}
z^{\bar 3}\over 2})),& \nonumber \\
f_2=&1+z^3z^{\bar 3}+
(z^2-{z^1z^3\over 2})(z^{\bar 2}-{z^{\bar 1}
z^{\bar 3}\over 2}).& \nonumber
\end{eqnarray}
Here $u$ and $u'$ are arbitrary constants which determine the value of the metric at the origin of the manifold. 
For this  K\"ahler potential the holomorphic functions $F^A(z)$ in (\ref{LieK}) are given by 
\begin{eqnarray}
F^1&=& -iuz^1, \hspace{5.5cm} F^{\ov 1}= 0, \nonumber \\
F^2&=& -i[u(z^2 +{1\over 2}z^1z^3) + u'(z^2 -{1\over 2}z^1z^3)], \ \ F^{\ov 2}= 0,\nonumber\\
F^3&=& -iu'z^2, \hspace{5.4cm} F^{\ov 3}= 0, \nonumber \\
F^{Q}&=& -i{u\over \sqrt 2}, \hspace{5.4cm} F^{Q'}= 
-i(\textstyle{1\over \sqrt 6}u + \textstyle{\sqrt{2\over 3}}u'). \nonumber
\end{eqnarray}
Putting these in (\ref{kill}) we find the Killing potentials as
\begin{eqnarray}
M^1&=&-{z^1\over f_1}u+{z^{\bar 3}(z^2-{1\over 2}z^1z^3)\over f_2}
u',\hspace{1.5cm} c.c., \nonumber\\
M^2&=&-{z^2+{1\over 2}z^1z^3\over f_1}u
-{z^2-{1\over 2}z^1z^3\over f_2}u',\ \hspace{0.5cm} c.c., \nonumber\\
M^3&=&-{z^{\bar 1}(z^2+{1\over 2}z^1z^3)\over f_1}u-
{z^3\over f_2}u', \hspace{1.5cm} c.c.,  \nonumber \\
M^{Q}&=&{z^1z^{\bar 1}-1\over \sqrt{2}f_1}u
+{1-2z^3z^{\bar 3}\over \sqrt{2}f_2}u'-
{1\over \sqrt{2}}u', \ \nonumber \\
M^{Q'}&=&\sqrt{2\over 3}{
(z^2+{1\over 2}z^1z^3)(z^{\bar 2}+{1\over 2}z^{\bar 1}z^{\bar 3})\over 
f_1}u-{1\over \sqrt{6}}u
-{1\over \sqrt{6}f_2}u'+{1\over \sqrt{6}}u', \nonumber
\end{eqnarray}
which satisfy the Lie-algebrae  (\ref{RR}) and 
$$M^AM^A=2(\sum_{i=1}^3M^iM^{\bar i})+(M^{Q})^2+(M^{Q'})^2=
{2\over 3}(u^2+uu'+u'^2).
$$
It is easy to  verify the fuzzy algebrae (\ref{eq 0}) and (\ref{eq 00}) with the $c_i = 0$ for $i\ge 3$ by rewriting the Killing potentials as (\ref{lll}) with the Darboux coordinates.

\vspace{1cm}

To conclude this letter we give an example for a non-commutative field theory  which 
exhibits  the fuzzy structure of the reducible K\"ahler coset space, but not through the quantum deformation discussed here. It is the boundary WZW model on a group manifold $G$\cite{11,10}, which is subject to the gluing condition
\begin{eqnarray}
g^{-1}\partial g  = -\bar \partial gg^{-1}    \label{glue}
\end{eqnarray}
at $z = \ov z$. This condition forces the string end $g|_{z=\bar z}$ to stay on a conjugacy class of an element $h \in G$
$$
{\cal C}(h) = \{lhl^{-1} \ | \ l\in G \}.
$$
It describes D-branes in the group manifold $G$. Without loosing the generality one may take $h$ as being made only of the Cartan generators, i.e., $ h = \exp (i\theta\cdot H)$.  Then there is an invariant subgroup $H$ such that 
$$
H = \{l\in G\  | \ lhl^{-1} = h \}.
$$
 In other words the gluing condition (\ref{glue}) breaks the group symmetry $G$ into $H$ with a given $h$. The conjugacy class ${\cal C}(h)$ describing D-branes is described by  coset spaces $G/H$. 
For $G = SU(3)$ we have  $SU(3)/SU(2)\otimes U(1)$ and $SU(3)/U(1)^2$ as non-trivial conjugate classes. The latter is an example of the reducible K\"ahler coset space which was of main interst in this letter. We have shown the fuzzy structure of the general K\"ahler coset space by studying quantum deformation of the coset space. But it can be shown  on D-branes of the boundary WZW model  in the semi-classical limit where the level $k$ of the model tends to $\infty$\cite{11}. In ref. \cite{12} it was analyzed at quantum level as well.

\vspace{1cm}

The fuzzy homogeneous space $G/H$ was also found in the Matrix Model
in ref. \cite{13}. It is described by using  
matrices of a finite representation of $G$, which satisfy  
the equation of motion of the Matrix model.
A  non-commutative field theory appears by considering
 fluctuations around this configuation. 
If the fuzzy homogeneous space $G/H$ is  k\"ahlerian, by taking the large limit of the matrix size, 
 one can see that the symplectic structure of the matrix description would
reduce to the one of the K\"ahler coset space  with the Darboux coordinates. In this limit the fuzzy algebrae are the same as the one 
we have  derived in this letter.

\vspace{1cm}

The fuzzy structure may be found also for the hyperk\"ahler manifold. For instance, the one with Eguchi-Hanson metric is characterized by the K\"ahler potential 
$$
K = \sqrt{1+\rho^4} -\ln {1+\sqrt{1+\rho^4} \over \rho^2}
$$
with $\rho^2 = |z^1|^2 + |z^2|^2$. Obviously the manifold has the isometry $U(2)$ given by the holomorphic Killing vectors
\begin{eqnarray}
R^{A\alpha} = i(T^Az)^\alpha,   \label{hk}
\end{eqnarray}
where $T^A$ are the $U(2)$ generators in the fundamental representation. Our arguments from (\ref{Kaform}) to (\ref{lll}) go through. Namely the Killing potentials take the forms
\begin{eqnarray}
iM^A = K_{,\alpha}R^{A\alpha}.  \label{M}
\end{eqnarray}
(Note that $F^A = 0$ in this case.) With the $\star$ product (\ref{Moy}) we find 
the fuzzy algebra (\ref{eq 0}). But the isometry $U(2)$ is merely realized linearly as (\ref{hk}). The arguments which lead us to the formula $M^AM^A = const$ for the generalized K\"ahler coset space are not applicable to this case. Indeed we have 
$$
M^AM^A = 2(1+\rho^4).
$$
Therefore the Killing potentials (\ref{M}) do not satisfy the fuzzy algebra in the form (\ref{eq 00}).

\vspace{2cm}
\noindent
{\Large\bf Acknowledgements}

T.M. would like to thank 
Y. Kitazawa and Y. Kimura for discussions. 
The work of S.A. was supported in part  by the Grant-in-Aid for Scientific Research No.
13135212.

\hspace{3cm}

\end{document}